%% file: main.tex
\documentclass[]{spie}  %>>> use for US letter paper
\usepackage{amsmath,amsfonts,amssymb}
\usepackage{graphicx}
\usepackage[colorlinks=true, allcolors=blue]{hyperref}

\title{Capabilities of a fibered imager on an extremely large telescope}

\author[a,b,c]{S. Vievard}
\author[d]{N. Cvetojevic}
\author[b]{E. Huby}
\author[b]{S. Lacour}
\author[e]{G. Martin}
\author[a,c,f,g]{O. Guyon}
\author[a]{J. Lozi}
\author[c]{T. Kotani}
\author[h]{N. Jovanovic}
\author[b]{G. Perrin}
\author[i]{F. Marchis}
\author[d]{O. Lai}
\author[b]{V. Lapeyrere}
\author[b]{D. Rouan}

\affil[a]{National Astronomical Observatory of Japan, Subaru Telescope, 650 North Aohoku Place,
Hilo, HI 96720, U.S.A.}
\affil[b]{Observatoire de Paris - LESIA, 5 Place Jules Janssen, 92190 Meudon, France}
\affil[c]{Astrobiology Center of NINS, 2-21-1, Osawa, Mitaka, Tokyo, 181-8588, Japan}
\affil[d]{Observatoire de la C\^ote d'Azur, 96 Boulevard de l'Observatoire, 06300 Nice, France}
\affil[e]{Univ. Grenoble Alpes, CNRS, IPAG, 38000 Grenoble, France}
\affil[f]{College of Optical Sciences, University of Arizona, Tucson, AZ 85721, U.S.A.}
\affil[g]{Jet Propulsion Laboratory, 4800 Oak Grove Drive, MS 183-901, Pasadena, CA 91109, U.S.A.}
\affil[h]{California Institute of Technology, 1200 E California Blvd, Pasadena, CA 91125, U.S.A.}
\affil[i]{Carl Sagan Center at the SETI Institute, 189 Bernardo Av., Mountain View, CA 94043, USA}

\authorinfo{Further author information: Sebastien Vievard: E-mail: vievard@naoj.org}

\begin{document} 
\maketitle

\begin{abstract}
FIRST, the Fibered Imager foR a Single Telescope instrument, is an ultra-high angular resolution spectro-imager, able to deliver calibrated images and measurements beyond the telescope diffraction limit, a regime that is out of reach for conventional AO imaging. 
FIRST achieves sensitivity and accuracy by coupling the full telescope to an array of single mode fibers. Interferometric fringes are spectrally dispersed and imaged on an EMCCD. An 18-Fiber FIRST setup is currently installed on the Subaru Coronographic Extreme Adaptive Optics instrument at Subaru telescope. It is being exploited for binary star system study. In the late 2020 it will be upgraded with delay lines and an active LiNb03 photonic beam-combining chip allowing phase modulation to nanometer accuracy at MHz. 
On-sky results at Subaru Telescope have demonstrated that, thanks to the ExAO system stabilizing the visible light wavefront, FIRST can acquire long exposure and operate on significantly fainter sources than previously possible. A similar approach on a larger telescope would therefore offer unique scientific opportunities for galactic (stellar physics, close companions) and extragalactic observations at ultra-high angular resolution. We also discuss potential design variations for nulling and high contrast imaging.

\end{abstract}

% Include a list of keywords after the abstract 

\input{1-Intro_first.tex}
\input{2-First_v1.tex}

\input{3-First_v2.tex}
\input{4-Capabilities_ELT.tex}

\section{COnclusion}

We presented the FIRST as a host module of the SCExAO instrument. FIRST already showed premises of its power by achieving a companion detection at half the diffraction limit of the 8-meter Subaru Telescope. More work on the acquired data and on the data reduction pipeline will surely bring new exciting science results in the near future. Moreover, new upgrades are already planned and will be implemented without disturbing the first version of the instrument on SCExAO. Finally, we discussed an estimation of the performance of FIRST on an ELT. FIRST, such as other photonics instruments would help to go beyond the diffraction limit of an ELT with high contrast abilities. Combined with the cutting-edge technology in terms of detectors or spectral dispersion, it would certainly allow to make new breakthroughs and discoveries.

\section{Acknowledgments}
The development of FIRST was supported by Centre National de la Recherche Scientifique CNRS (Grant ERC LITHIUM - STG - 639248). The development of SCExAO was supported by the Japan Society for the Promotion of Science (Grant-in-Aid for Research \#23340051, \#26220704, \#23103002, \#19H00703 \& \#19H00695), the Astrobiology Center of the National Institutes of Natural Sciences, Japan, the Mt Cuba Foundation and the director's contingency fund at Subaru Telescope. The authors wish to recognize and acknowledge the very significant cultural role and reverence that the summit of Maunakea has always had within the indigenous Hawaiian community. We are most fortunate to have the opportunity to conduct observations from this mountain.

% References
\bibliography{main} % bibliography data in report.bib
\bibliographystyle{spiebib} % makes bibtex use spiebib.bst

\end{document}

%% file: 1-Intro_first.tex
\section{INTRODUCTION}
\label{sec:intro}  % \label{} allows reference to this section

One of the key challenges in astronomy is the detection and characterization of faint companions such as exoplanets close to their host star. Achieving this requires both high performances in terms of angular resolution and dynamic range. Even though very large telescopes (8-meter class) were built and Extremely Large Telescopes (ELTs, 30-meter class) are on the horizon, reaching the desired performances is quite challenging especially for those ground based telescopes subject to turbulent atmosphere. 

To reach the diffraction limit of a telescope, several techniques were developed. Among them, the Adaptive Optics (AO) technique~\cite{rousset1990first} consists in using a deformable mirror to correct for the turbulence-induce aberrations on the incoming wavefront, allowing to restore the diffraction limit. However, because this correction is not perfect, residual speckle noise still limits the achievable dynamic range.

Other techniques were investigated. Speckle interferometry~\cite{labeyrie1970attainment} , based on the post processing of short exposure images, also allows to reach the diffraction limit of the telescope but still limits the dynamic range to a few 100. Aperture masking consists in placing a non-redundant mask in the pupil of the telescope and recombine the coherent light from each sub-aperture. This allows to retrieve spatial information at the highest spatial frequency of the telescope with contrast ratios down to $10^{-3}$~\cite{lacour2011sparse} . However it necessitate to sacrifice a large percentage of the full pupil because of the need of non-redundancy for the mask. In addition, speckle noise remains across each sub-aperture.

The method that can overcome the limitations of the aperture masking technique is the pupil remapping combined with the use of single-mode fibers\cite{perrin2006high} . On one hand, the pupil remapping allows one to use the whole aperture. On the other hand, the spatial filtering offered by the single-mode fibers allows to completely remove speckle noise over each sub-aperture. An instrument with such features is called FIRST (Fibered Imager foR a Single Telescope) and is currently installed on the 8-meter Subaru Telescope as part of the Subaru Coronographic Extreme Adaptive Optics~\cite{2015PASP..127..890J} (SCExAO) instrument. During its early development, FIRST revealed its power on the 3-meters Lick observatory performing spectroscopy analysis of the Capella binary system at the diffraction limit~\cite{huby2012first,huby2013first} . In this paper we present the integration and principle of FIRST as a module of SCExAO. We then present new upgrades for a version 2 (FIRST v2) of the instrument. Finally we discuss the interest of implementing FIRST, and more generally the interest of photonics instrument, on an ELT. 

%% file: 2-First_v1.tex
\section{FIRST: pupil-remapping interferometry at the Subaru Telescope}

\subsection{FIRST, a module on SCExAO}
\begin{figure}[!h]
    \centering
    \begin{tabular}{cc}
        \includegraphics[width=0.47\linewidth]{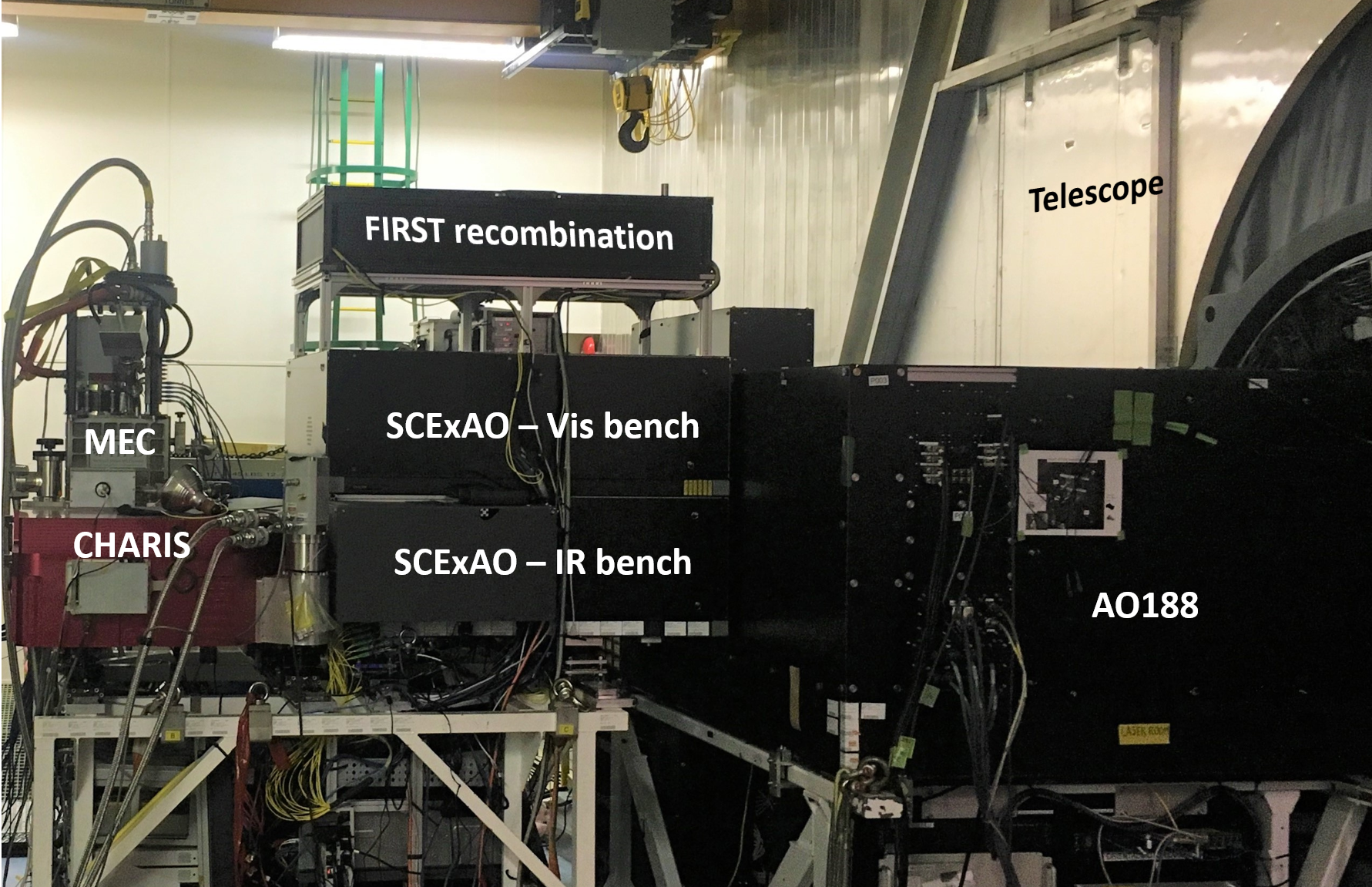} & \includegraphics[width=0.47\linewidth]{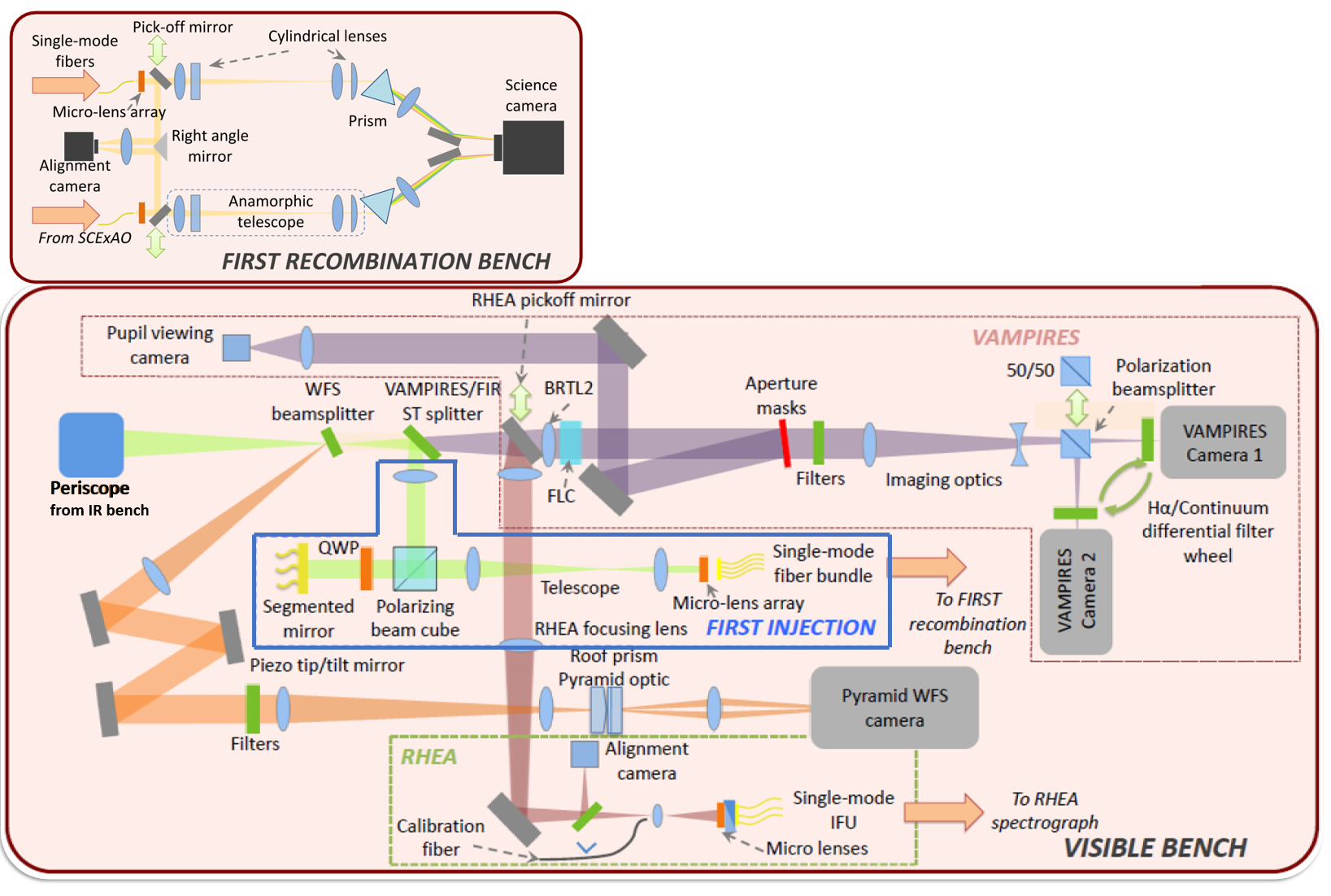} \\
    \end{tabular}{}
    \caption{Left: SCExAO on the Infra-Red (IR) Nasmyth platfrom sits after AO188. SCExAO is composed of two benches on the top of the other, plus the FIRST recombination bench. SCExAO delivers light to CHARIS~\cite{Currie_2018} and MEC~\cite{walter2018mec} detectors. Right: A schematic of the optic path in the Visible bench and in the FIRST recombination bench.}
    \label{fig:scexao_irnas}
\end{figure}

SCExAO is an instrument that is located on the Infra-Red (IR) Nasmyth platform of the Subaru telescope (see Fig.~\ref{fig:scexao_irnas}). It is fed by AO188~\cite{minowa2010performance}, a 188-element Curvature sensor adaptive optic system that delivers a first stage of wavefront correction. SCExAO main wavefront sensor is a pyramid wavefront sensor (PyWFS) in the visible (around $850$nm - see \textit{Lozi et al.}~\cite{Lozi_2019} for more details) delivering a wavefront quality over 80\% Strehl (in H-band), critical for high contrast imaging. SCExAO hosts a lot of different modules, and is divided over two benches: an IR and a Visible (Vis) bench. A dichroic filter on the IR bench and sends light (wavelengths $<900$nm) on the Vis bench through a periscope (see Fig.~\ref{fig:scexao_irnas}, right).

{}

The PyWFS usually operates between $850$ and $950$nm thanks to a pickoff filter wheel. Then the light can be split between VAMPIRES~\cite{norris2015vampires} and FIRST using a set of different filters on a filter wheel. All the light after the pyramid pickoff can also be sent either to FIRST with a mirror, or to VAMPIRES with an open slot.

\newpage
\subsection{Principle and setup}
\begin{figure}[!h]
    \centering
    \includegraphics[width=\linewidth]{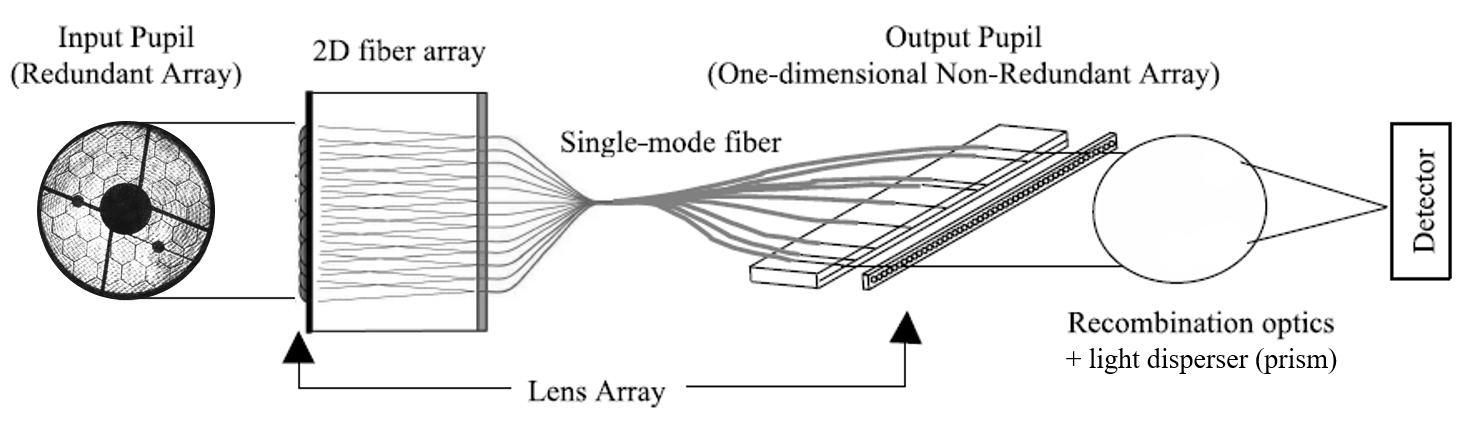}
    \caption{Schematic diagram~\cite{kotani2009pupil} showing the principle of the FIRST instrument at the Subaru Telescope. The pupil of the Subaru Telescope is divided into several single-mode fibers thanks to a 2D micro-lens array. Light from the different fibers are then set into a one-dimensional non-redundant array, recombined and spectrally dispersed.  }
    \label{fig:firstv1_ppe}
\end{figure}

\begin{figure}[!h]
    \centering
    \includegraphics[width=0.7\linewidth]{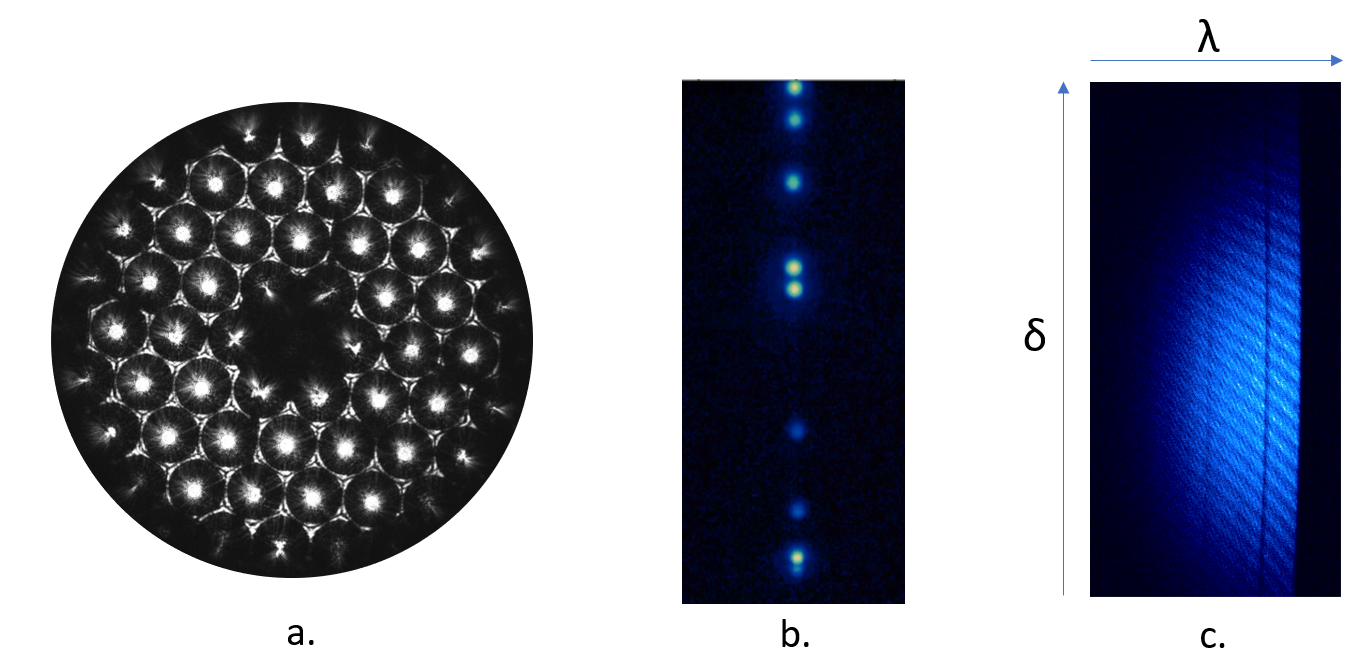}
    \caption{a. Image of the micro-lens array conjugated with the Iris AO MEMs. b. One-dimesion non-redundant array: output of the V-groove from one set of 9 fibers. c. Recombination of one set of 9 fibers spectrally dispersed.}
    \label{fig:firstv1_ppe2}
\end{figure}

FIRST combines the two techniques of pupil remapping and spatial filtering with single-mode fibers. Fig.~\ref{fig:firstv1_ppe} shows the principle of the instrument. The pupil of the telescope is divided into 18 different single mode fibers. The input pupil, redundant, is injected in the single-mode fibers with a 2D micro-lens array. The optimization of the injection is performed by an Iris AO segmented MEMS. Each segment is conjugated with a micro-lens (see Fig.~\ref{fig:firstv1_ppe2}a.) which focuses the beam into a single-mode fiber. A fiber bundle contains in total 36 single-mode fibers, that are then each connected at the other end to extension fibers that run to the recombination bench (see Fig.~\ref{fig:scexao_irnas}). On the recombination bench, the extension fibers are connected to single-mode fibers that end in a V-groove. The 18 used fibers form two sets of 9 fibers. The output of the two sets is 2 one-dimensional non redundant arrays each formed in a V-groove. Therefore, for each set of fibers, the N redundant baseline from the input pupil are now N non-redundant baselines corresponding to $N(N-1)/2$ different spatial frequencies. The outputs of each fibers are collimated (see Fig.~\ref{fig:firstv1_ppe2}b.) then recombined and spectrally dispersed with an equilateral SF2-prism. In order to increase the spectral resolution, an afocal anamorphic system, consisting in cylindrical lenses, stretches the beam in the dispersion direction and compresses it in the orthogonal direction. As shown on Fig~\ref{fig:firstv1_ppe2}c. the acquired fringe pattern contains information in terms of Optical Path Difference (OPD, phase, $\delta$) and in terms of spectrum (wavelength, $\lambda$). The spectral resolution obtained is empirically around $300$ @$700$nm.

{}

{}

\subsection{Data reduction}\label{firstv1:data}

Typical FIRST images are two sets of fringes (one is shown on Fig~\ref{fig:firstv1_ppe2}c). The two sets are reduced separately, but rely on the same fringe analysis technique. After substraction of the dark, wavelength calibration and data curvature (caused by our optics) correction, we do a fringe fitting in the focal plane using the P2VM~\cite{millour2004data} method in each wavelength channel (each column of the image). This allows to compute closure phases for different baselines combinations. One nice feature of the closure phase measurement is that it allows to cancel differential phase errors between sub-apertures that can remain even after single-mode fiber filtering, Fitting a model of closure phase on the estimated closure phases allows to access spatial information about the observed object below the diffraction limit of the telescope. 
A new user-friendly data reduction pipeline is currently under development and should be completed by the beginning of 2020 to broaden the number of FIRST users.

\subsection{Science with FIRST}
Typical targets of interest for FIRST are close binary systems or surface of large resolved stars. An ultimate goal for this concept would be the direct detection and characterization of an exoplanet.
At the Subaru Telescope, FIRST already showed that it could detect a companion below the diffraction limit (18 mas @700nm). Indeed, on the Alpha Equ binary system, a companion was detected with at a separation of around 10mas, or almost half of the telescope diffraction limit. These data are still under analysis.

%% file: 3-First_v2.tex
\section{FIRSTv2: photonics upgrades}

\begin{figure}[!h]
    \centering
    \includegraphics[width=\linewidth]{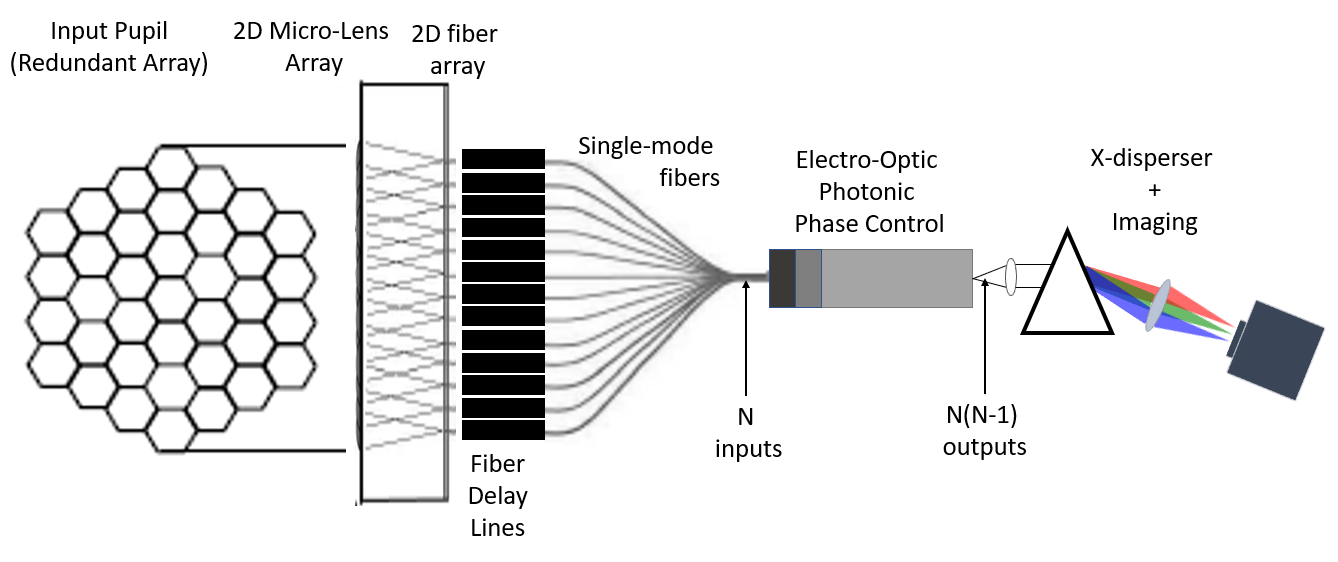}
    \caption{Schematic diagram showing the principle of FIRSTv2. Upgrades compared to FIRSTv1 are the fiber delay lines, the electro-optic photonic phase controller-recombiner and the camera.}
    \label{fig:firstv2_ppe}
\end{figure}

While FIRST can still take data on-sky, upgrades are in preparation. One of the advantages to have two sets of 9 fibers is that one of them can be upgraded while the other is still in use. Fig.~\ref{fig:firstv2_ppe} shows the schematic diagram of FIRST v2, with three major upgrades.

{}

\subsection{Fiber path length matching}
\begin{figure}[!ht]
    \centering
    \includegraphics[width=\linewidth]{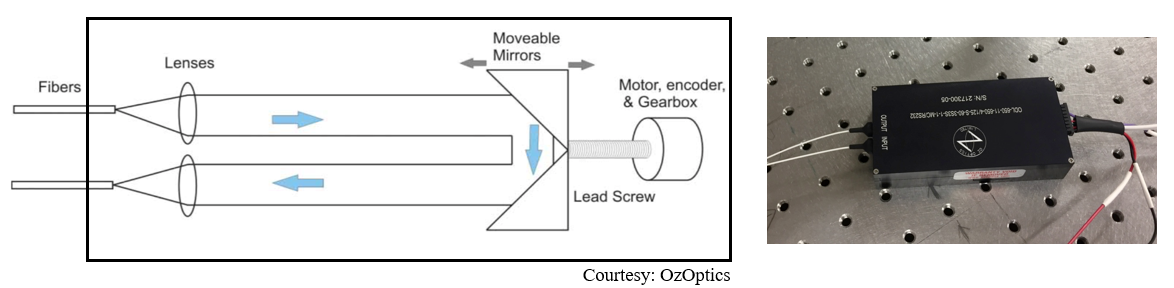}
    \caption{Left: Optical path inside the delay line. The input beam reflects on a movable mirror and is then re-injected in the output fiber. Right: Picture of a delay line.}
    \label{fig:Delay_lines}
\end{figure}

Current single-mode fiber extensions on FIRST were all manufactured together and then polished by hand to match the path length. If/whenever one breaks it is then time consuming and complicated to make a new one and match the path length with the others. To prevent this, and to have a better accuracy on the path length matching of the current fibers, we want to implement fiber delay lines (see Fig.~\ref{fig:Delay_lines}) between the fibers from the fiber bundle and the extension fibers.

{}

Fig.~\ref{fig:Delay_lines} shows on the left the optic path inside the delay line. The light from the input fiber is collimated, bounces on two movable mirrors and is then re-injected in the output fiber. The mirrors sit on a lead screw that is linked to a motor. The total OPD range is of $100mm$, the OPD resolution is $0.4134\mu m$ and the OPD accuracy is $1$ to $2\mu m$. 

\subsection{Active photonic chip}
\begin{figure}[!ht]
    \centering
    \includegraphics[width=\linewidth]{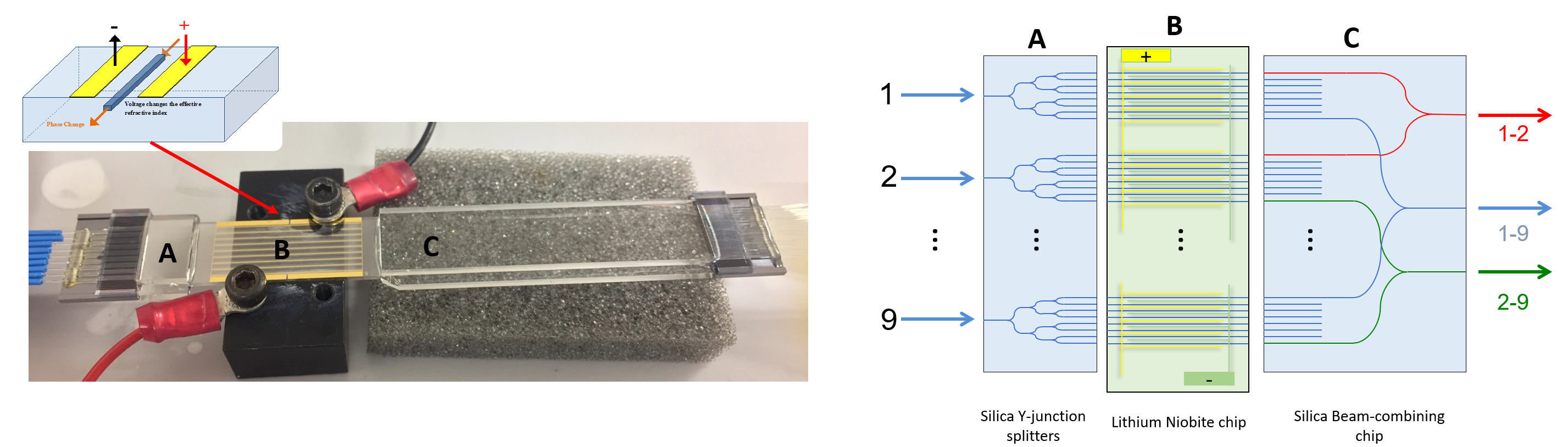}
    \caption{New integrated optic active chip composed by 3 elements: the Silica Y-junctions that split each input, the Lithium Niobate chip that allows a phase control within each waveguides and the Silica beam-combining chip to form recombine the different pairs.}
    \label{fig:firstv2_iochip}
\end{figure}

The second upgrade is the integration of an integrated optic active chip. Fig.~\ref{fig:firstv2_iochip} shows the three different components of this chip:
\begin{itemize}
    \item \textbf{A: The silica Y-junction splitters} split each of the 9 input waveguides in 8. Meaning there are 9 inputs and 72 outputs.
    \item \textbf{B: The Lithium Niobate chip} takes in the 72 straight waveguides. As shown Fig.~\ref{fig:firstv2_iochip} a potential difference can be applied across the waveguides. The consequence of this is a change of the Lithium Niobate refractive index, modifying the phase in the waveguides. This allows to control the phase delay inside each waveguide at high speed (MHz).
    \item \textbf{C: The Silica beam-combining chip} takes in the 72 waveguides and recombines them two by two. Doing so, we have 36 outputs that are the 36 different baselines (similar to current FIRST version). The difference here is that instead of being all superimposed, the lights from each baseline are well separated at the output of the combining chip. A phase ramp is then applied to obtain the OPD information.
\end{itemize}{}

{}

\subsection{New output data}
\begin{figure}[!ht]
    \centering
    \includegraphics[width=0.4\linewidth]{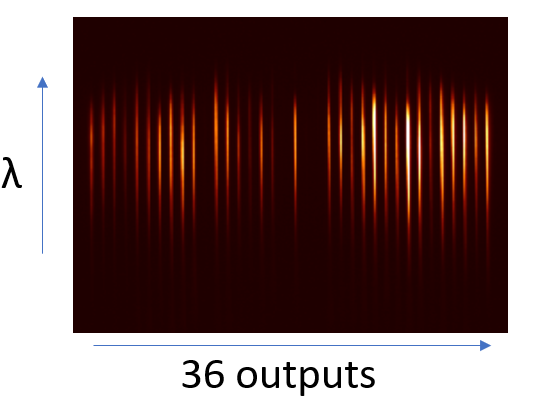}
    \caption{Data output for FIRST v2. Each of the 36 outputs of the beam-combining chip are dispersed with a prism, and are all separated.}
    \label{fig:firstv2_cam}
\end{figure}

As mentioned in Section~\ref{firstv1:data}, the current data in FIRST is the superimposition of all the interference patterns between the different baselines, spectrally dispersed. Each image then contains spectral and OPD information. With FIRST v2, the output data is little different. A typical image on the camera of the output of the beam-combining chip after spectral dispersion is shown on Fig.~\ref{fig:firstv2_cam}.

{}

In this case, one image only gives a spectral information and an intensity. To retrieve the OPD information, one needs to modulate the phase in each waveguide and record series of images: the fringes are temporally encoded. To do so a voltage ramp is applied on the chip B (Fig.~\ref{fig:firstv2_iochip}) to temporally scan the OPD For the data reduction, a similar fringe fitting using the V2PM method is used.

All these upgrades will be brought to the Subaru Telescope in late 2020 or early 2021, and could potentially increase the stability and sensitivity of the instrument.

%% file: 4-Capabilities_ELT.tex
\section{Capabilities of a photonic imager on an Extremely Large Telescope}

We saw how the FIRST instrument concept could be powerful on an 8m telescope like Subaru, enabling detection and spectroscopy below the limit of the telescope. When the concept was born, it was already envisioned~\cite{perrin2006high} for 30-meter class telescopes. We would like to bring forward again this vision at the end of this paper, presenting different use of FIRST on a segmented ELT and trying to estimate the performance that could be expected.

\subsection{FIRST-like instrument on an ELT}
The spatial resolution of a 30-meter class telescope is around $5$mas @$700$nm. According to our experience at the Subaru Telescope, we saw that FIRST was able to detect a companion at roughly half of the diffraction limit. This would mean that FIRST on a 30-meter class telescope could potentially be able to detect and characterize companions at around $2.5$mas from the host star. 

In terms of achievable contrast \textit{B. Norris and al.}~\cite{norris2014high} showed that in the case of a pupil-remapping interferometer, the contrast ratio detection at $1\sigma$ was directly proportional to the on-sky closure-phase stability. The latest not only relies on the stability of the instrument itself, but also on the stability of the adaptive optics system. Since an high contrast imager on an ELT would most likely have an extreme adaptive optics system, it is easy to infer that the ExAO-corrected phase error would be a few dozens of nm. The stability of the closure phase in-lab is around $0.01\deg$ from experience. We can then estimate thanks to \textit{B. Norris and al.}~\cite{norris2014high} equations (4) and (5) that the achievable contrast of FIRST on an ELT would be of the order of $10^{-6}$. 

\begin{figure}[!h]
    \centering
    \includegraphics[width=0.45\linewidth]{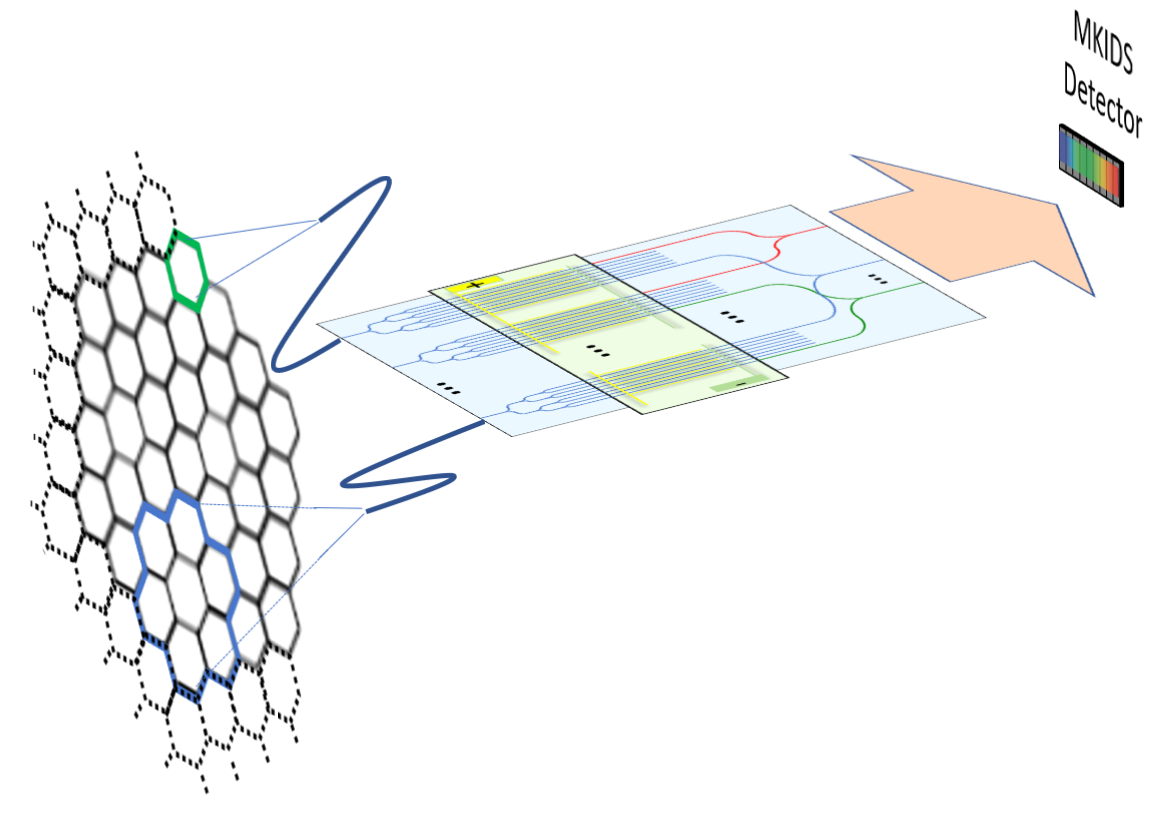}
    \caption{FIRST concept on a large multi-aperture telescope
(small portion of the segmented mirror shown here). Light from different amount of segments can be injected in a single-mode fiber, changing the sensitivity of the instrument. The light from the fibers could then go through an integrated optics active chip, whose output could be imaged on a read-out-noise-free MKIDS detector.}
    \label{fig:first_elt}
\end{figure}

Finally, the sensitivity of the instrument scales with the amount of light injected in the single-mode fibers hence the size of the sub-aperture. Fig.~\ref{fig:first_elt} shows two different setups that can be imagined for FIRST on an ELT. If only one or seven segments per sub-aperture are shown here, in practice we would want more sub-apertures per fiber to maximize the sensitivity. Another way to increase the sensitivity of the instrument would be to use an MKIDS~\cite{walter2018mec} detector or on-chip MKIDS-like detector~\cite{cheng2019broadband} allowing to have a spectrometer directly on the chip with no read-out-noise.
%If the size of the sub-aperture was 1 segment, it would be quite similar to the size of the sub-apertures at the Subaru Telescope where we empirically found that the magnitude limit of our instrument was around $6.5$. In similar conditions, we can say that it would be the same for a configuration with one segment on an ELT. An extrapolation allows us to estimate that if the sub-apertures were composed by 7 segments, the magnitude limit would be around $11.5$.

{}

\subsection{Other possibilities with photonics}
An often overlooked benefit of the ELT generation is that the huge telescope diameter brings single-aperture interferometric techniques (FIRST, GLINT~\cite{lagadec2018glint} , aperture masking) closer to the realm of long-baseline interferometry (GRAVITY~\cite{eisenhauer2007gravity} , OHANA~\cite{perrin2000fibered}). In fact, many techniques and technologies developed for long-baseline interferometry can be directly applied to ELTs, with minimal modification (as we discussed in the previous section in the particular case of FIRST).

\begin{figure}[!h]
    \centering
    \includegraphics[width=0.9\linewidth]{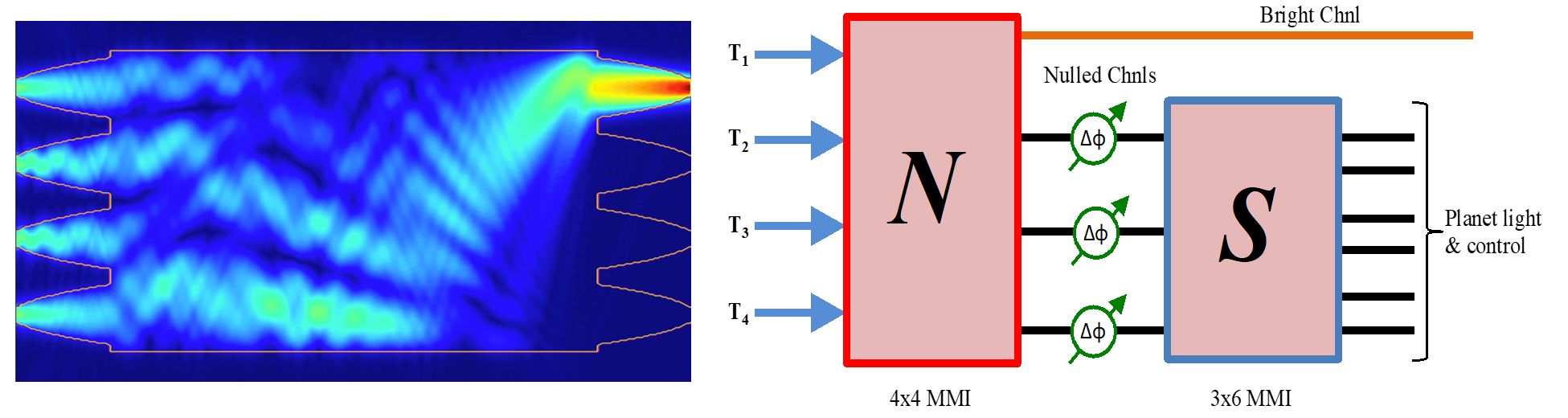}
    \caption{Left: Beam propagation simulation of a MultiMode interference coupler (MMI) corresponding to the $4\times 4$ MMI
    (\textbf{N}). The four input beams are interfered such that all starlight is sent to the bright output port. Right: The Kernel-nuller
    concept composed of $2$ MMIs separated by modulated phase shifters used to eliminate background fluctuations. The $3$ 'dark' ports
    feed the kernel-scrambler (\textbf{S}) matrix which produces the kernel-null observables. }
    \label{fig:kernel}
\end{figure}

More interestingly, ELTs bring a hybrid space perfect for new technologies. One such technology is on-chip kernel nulling~\cite{martinache2018kernel,goldsmith2017improving} . In this concept (see Fig.~\ref{fig:kernel}), photonic circuitry consisting of cascaded multimode interference couplers (MMIs) interferes the light in a particular way to not only null the starlight (destructively interfere on-axis starlight that causes photon noise), but additionally interfere the light in a particular way such that the resulting flux outputs form Kernel observables. These Kernels have the same properties as closure phase, and are extremely robust against second order phase error from the atmosphere. They also are a way to increase the achievable contrast compared to previous section. A simple analogy would be to combine the detection contrasts of a coronagraph with the astrometry precision of GRAVITY and the spectral capability of IRD~\cite{kotani2018infrared} . 

{}

%% file: main.bbl
\begin{thebibliography}{10}

\bibitem{rousset1990first}
Rousset, G., Fontanella, J., Kern, P., Gigan, P., and Rigaut, F., ``First
  diffraction-limited astronomical images with adaptive optics,'' {\em
  Astronomy and Astrophysics}~{\bf 230},  L29--L32 (1990).

\bibitem{labeyrie1970attainment}
Labeyrie, A., ``Attainment of diffraction limited resolution in large
  telescopes by fourier analysing speckle patterns in star images,'' {\em
  Astron. Astrophys.}~{\bf 6}(1),  85--87 (1970).

\bibitem{lacour2011sparse}
Lacour, S., Tuthill, P., Amico, P., Ireland, M., Ehrenreich, D., Huelamo, N.,
  and Lagrange, A.-M., ``Sparse aperture masking at the vlt-i. faint companion
  detection limits for the two debris disk stars hd 92945 and hd 141569,'' {\em
  Astronomy \& Astrophysics}~{\bf 532},  A72 (2011).

\bibitem{perrin2006high}
Perrin, G., Lacour, S., Woillez, J., and Thi{\'e}baut, E., ``High dynamic range
  imaging by pupil single-mode filtering and remapping,'' {\em Monthly Notices
  of the Royal Astronomical Society}~{\bf 373}(2),  747--751 (2006).

\bibitem{2015PASP..127..890J}
{Jovanovic}, N., {Martinache}, F., {Guyon}, O., {Clergeon}, C., {Singh}, G.,
  {Kudo}, T., {Garrel}, V., {Newman}, K., {Doughty}, D., {Lozi}, J., {Males},
  J., {Minowa}, Y., {Hayano}, Y., {Takato}, N., {Morino}, J., {Kuhn}, J.,
  {Serabyn}, E., {Norris}, B., {Tuthill}, P., {Schworer}, G., {Stewart}, P.,
  {Close}, L., {Huby}, E., {Perrin}, G., {Lacour}, S., {Gauchet}, L.,
  {Vievard}, S., {Murakami}, N., {Oshiyama}, F., {Baba}, N., {Matsuo}, T.,
  {Nishikawa}, J., {Tamura}, M., {Lai}, O., {Marchis}, F., {Duchene}, G.,
  {Kotani}, T., and {Woillez}, J., ``{The Subaru Coronagraphic Extreme Adaptive
  Optics System: Enabling High-Contrast Imaging on Solar-System Scales},'' {\em
  Publications of the Astronomical Society of the Pacific}~{\bf 127},  890
  (Sept. 2015).

\bibitem{huby2012first}
Huby, E., Perrin, G., Marchis, F., Lacour, S., Kotani, T., Duch{\^e}ne, G.,
  Choquet, E., Gates, E., Woillez, J., Lai, O., et~al., ``First, a fibered
  aperture masking instrument-i. first on-sky test results,'' {\em Astronomy \&
  Astrophysics}~{\bf 541},  A55 (2012).

\bibitem{huby2013first}
Huby, E., Duch{\^e}ne, G., Marchis, F., Lacour, S., Perrin, G., Kotani, T.,
  Choquet, {\'E}., Gates, E., Lai, O., and Allard, F., ``First, a fibered
  aperture masking instrument-ii. spectroscopy of the capella binary system at
  the diffraction limit,'' {\em Astronomy \& Astrophysics}~{\bf 560},  A113
  (2013).

\bibitem{minowa2010performance}
Minowa, Y., Hayano, Y., Oya, S., Watanabe, M., Hattori, M., Guyon, O., Egner,
  S., Saito, Y., Ito, M., Takami, H., et~al., ``Performance of subaru adaptive
  optics system ao188,'' in [{\em Adaptive Optics Systems
  II}{\nolinebreak\hspace{0.1em}]},   {\bf 7736},  77363N, International
  Society for Optics and Photonics (2010).

\bibitem{Lozi_2019}
Lozi, J., Jovanovic, N., Guyon, O., Chun, M., Jacobson, S., Goebel, S., and
  Martinache, F., ``Visible and near-infrared laboratory demonstration of a
  simplified pyramid wavefront sensor,'' {\em Publications of the Astronomical
  Society of the Pacific}~{\bf 131},  044503 (mar 2019).

\bibitem{Currie_2018}
Currie, T., Brandt, T.~D., Uyama, T., Nielsen, E.~L., Blunt, S., Guyon, O.,
  Tamura, M., Marois, C., Mede, K., Kuzuhara, M., Groff, T.~D., Jovanovic, N.,
  Kasdin, N.~J., Lozi, J., Hodapp, K., Chilcote, J., Carson, J., Martinache,
  F., Goebel, S., Grady, C., McElwain, M., Akiyama, E., Asensio-Torres, R.,
  Hayashi, M., Janson, M., Knapp, G.~R., Kwon, J., Nishikawa, J., Oh, D.,
  Schlieder, J., Serabyn, E., Sitko, M., and Skaf, N., ``{SCExAO}/{CHARIS}
  near-infrared direct imaging, spectroscopy, and forward-modeling of kappa and
  b: A likely young, low-gravity superjovian companion,'' {\em The Astronomical
  Journal}~{\bf 156},  291 (nov 2018).

\bibitem{walter2018mec}
Walter, A., Mazin, B.~B., Bockstiegel, C., Fruitwala, N., Szypryt, P.,
  Lipartito, I., Meeker, S., Zobrist, N., Collura, G., Coiffard, G., et~al.,
  ``Mec: the mkid exoplanet camera for high contrast astronomy at subaru
  (conference presentation),'' in [{\em Ground-based and Airborne
  Instrumentation for Astronomy VII}{\nolinebreak\hspace{0.1em}]},   {\bf
  10702},  107020V, International Society for Optics and Photonics (2018).

\bibitem{norris2015vampires}
Norris, B., Schworer, G., Tuthill, P., Jovanovic, N., Guyon, O., Stewart, P.,
  and Martinache, F., ``The vampires instrument: imaging the innermost regions
  of protoplanetary discs with polarimetric interferometry,'' {\em Monthly
  Notices of the Royal Astronomical Society}~{\bf 447}(3),  2894--2906 (2015).

\bibitem{kotani2009pupil}
Kotani, T., Lacour, S., Perrin, G., Robertson, G., and Tuthill, P., ``Pupil
  remapping for high contrast astronomy: results from an optical testbed,''
  {\em Optics Express}~{\bf 17}(3),  1925--1934 (2009).

\bibitem{millour2004data}
Millour, F., Tatulli, E., Chelli, A.~E., Duvert, G., Zins, G., Acke, B., and
  Malbet, F., ``Data reduction for the amber instrument,'' in [{\em New
  Frontiers in Stellar Interferometry}{\nolinebreak\hspace{0.1em}]},   {\bf
  5491},  1222--1230, International Society for Optics and Photonics (2004).

\bibitem{norris2014high}
Norris, B., Cvetojevic, N., Gross, S., Jovanovic, N., Stewart, P.~N., Charles,
  N., Lawrence, J.~S., Withford, M.~J., and Tuthill, P., ``High-performance 3d
  waveguide architecture for astronomical pupil-remapping interferometry,''
  {\em Optics express}~{\bf 22}(15),  18335--18353 (2014).

\bibitem{cheng2019broadband}
Cheng, R., Zou, C.-L., Guo, X., Wang, S., Han, X., and Tang, H.~X., ``Broadband
  on-chip single-photon spectrometer,'' {\em Nature communications}~{\bf
  10}(1),  1--7 (2019).

\bibitem{lagadec2018glint}
Lagadec, T., Norris, B., Gross, S., Arriola, A., Gretzinger, T., Cvetojevic,
  N., Lawrence, J., Withford, M., and Tuthill, P., ``Glint south: A photonic
  nulling interferometer pathfinder at the anglo-australian telescope for high
  contrast imaging of substellar companions,'' in [{\em Optical and Infrared
  Interferometry and Imaging VI}{\nolinebreak\hspace{0.1em}]},   {\bf 10701},
  107010V, International Society for Optics and Photonics (2018).

\bibitem{eisenhauer2007gravity}
Eisenhauer, F., Perrin, G., Rabien, S., Eckart, A., Lena, P., Genzel, R.,
  Abuter, R., Paumard, T., and Brandner, W., ``Gravity: the ao-assisted,
  two-object beam-combiner instrument for the vlti,'' in [{\em The Power of
  Optical/IR Interferometry: Recent Scientific Results and 2nd Generation
  Instrumentation}{\nolinebreak\hspace{0.1em}]},   431--444, Springer (2007).

\bibitem{perrin2000fibered}
Perrin, G.~S., Lai, O., Lena, P.~J., and du~Foresto, V.~C., ``Fibered large
  interferometer on top of mauna kea: Ohana, the optical hawaiian array for
  nanoradian astronomy,'' in [{\em Interferometry in Optical
  Astronomy}{\nolinebreak\hspace{0.1em}]},   {\bf 4006},  708--714,
  International Society for Optics and Photonics (2000).

\bibitem{martinache2018kernel}
Martinache, F. and Ireland, M.~J., ``Kernel-nulling for a robust direct
  interferometric detection of extrasolar planets,'' {\em arXiv preprint
  arXiv:1802.06252}  (2018).

\bibitem{goldsmith2017improving}
Goldsmith, H.-D.~K., Ireland, M., Ma, P., Cvetojevic, N., and Madden, S.,
  ``Improving the extinction bandwidth of mmi chalcogenide photonic chip based
  mir nulling interferometers,'' {\em Optics express}~{\bf 25}(14),
  16813--16824 (2017).

\bibitem{kotani2018infrared}
Kotani, T., Tamura, M., Nishikawa, J., Ueda, A., Kuzuhara, M., Omiya, M.,
  Hashimoto, J., Ishizuka, M., Hirano, T., Suto, H., et~al., ``The infrared
  doppler (ird) instrument for the subaru telescope: instrument description and
  commissioning results,'' in [{\em Ground-based and Airborne Instrumentation
  for Astronomy VII}{\nolinebreak\hspace{0.1em}]},   {\bf 10702},  1070211,
  International Society for Optics and Photonics (2018).

\end{thebibliography}
